\documentclass[twocolumn,showpacs,aps,prl,superscriptaddress,letterpaper]{revtex4}

\usepackage{graphicx}
\usepackage{dcolumn}
\usepackage{amsmath}
\usepackage{epsfig}
\usepackage{multirow}

\long\def\inst#1{\par\nobreak\kern 4pt\nobreak
    {\itshape #1}\par\vskip 10pt plus 3pt minus 3pt}
\RequirePackage{xspace}
\usepackage{relsize}

\def\babar{\mbox{\slshape B\kern-0.1em{\smaller A}\kern-0.1em
    B\kern-0.1em{\smaller A\kern-0.2em R}}}
\def\Abar    {\kern 0.18em\overline{\kern -0.18em A}{}\xspace}
\def\Kbar    {\kern 0.18em\overline{\kern -0.18em K}{}\xspace}
\def\Dbar    {\kern 0.18em\overline{\kern -0.18em D}{}\xspace}
\def\Bbar    {\kern 0.18em\overline{\kern -0.18em B}{}\xspace}

\def\BB      {\ensuremath{B\Bbar}\xspace} 
\def\Bz      {\ensuremath{B^0}\xspace}
\def\Bzb     {\ensuremath{\Bbar^0}\xspace}
\def\BzBzb   {\ensuremath{\Bz {\kern -0.16em \Bzb}}\xspace}
\def\Bu      {\ensuremath{B^+}\xspace}
\def\Bub     {\ensuremath{B^-}\xspace}

\def\BpBm    {\ensuremath{\Bu {\kern -0.16em \Bub}}\xspace}

\newcommand{\optbar}[1]{\shortstack{{\tiny (\rule[.4ex]{1em}{.1mm})}
  \\ [-.7ex] $#1$}}
\def\BorBbar    {\kern 0.18em\optbar{\kern -0.18em B}{}\xspace}
\def\DorDbar    {\kern 0.18em\optbar{\kern -0.18em D}{}\xspace}
\def\KorKbar    {\kern 0.18em\optbar{\kern -0.18em K}{}\xspace}

\def\pep2{PEP-II}
\mathchardef\Upsilon="7107
\def\Y#1S{\ensuremath{\Upsilon{(#1S)}}\xspace}

\def\FourS {\Y4S}

\newcommand{\BABARPubYear}     {07}
\newcommand{\BABARPubNumber}  {026}

\newcommand{\SLACPubNumber} {12462}

\begin{document}

\begin{flushleft}
\babar-PUB-\BABARPubYear/\BABARPubNumber\\
SLAC-PUB-\SLACPubNumber
\\[10mm]
\end{flushleft}

\title{
\large \bfseries \boldmath
Amplitude Analysis of the $B^\pm\to\varphi K^{*}(892)^\pm$ Decay 
}

%
\author{B.~Aubert}
\author{M.~Bona}
\author{D.~Boutigny}
\author{Y.~Karyotakis}
\author{J.~P.~Lees}
\author{V.~Poireau}
\author{X.~Prudent}
\author{V.~Tisserand}
\author{A.~Zghiche}
\affiliation{Laboratoire de Physique des Particules, IN2P3/CNRS et Universit\'e de Savoie, F-74941 Annecy-Le-Vieux, France }
\author{J.~Garra~Tico}
\author{E.~Grauges}
\affiliation{Universitat de Barcelona, Facultat de Fisica, Departament ECM, E-08028 Barcelona, Spain }
% ...
\collaboration{The \babar\ Collaboration}
\noaffiliation

\date{May 12, 2007}

\begin{abstract}
We perform an amplitude analysis of $B^\pm\to\varphi(1020)K^{*}(892)^\pm$ decay 
with a sample of about 384 million $\BB$ pairs recorded with the $\babar$ 
detector. Overall, twelve parameters are measured, including the fractions
of longitudinal ${f_L}$ and parity-odd transverse ${f_\perp}$ amplitudes,
branching fraction, strong phases, and six parameters sensitive to $C\!P$-violation.
We use the dependence on the $K\pi$ invariant mass of the interference between 
the $J^P=1^-$ and $0^+$ $K\pi$ components to resolve the discrete 
ambiguity in the determination of the strong and weak phases.
Our measurements of ${f_L}=0.49\pm{0.05}\pm 0.03$, ${f_\perp}=0.21\pm 0.05\pm 0.02$,
and the strong phases point to the presence of a substantial 
helicity-plus amplitude from a presently unknown source.
\end{abstract}

\pacs{13.25.Hw, 13.88.+e, 11.30.Er}

\maketitle


The polarization anomaly in vector-vector charmless hadronic $B$-meson 
decays~\cite{babar:vv, belle:phikst, belle:rhokst, 
babar:rhokst, babar:vt, bvvreview2006, pdg2006}
motivates a revision in our understanding of the effective 
flavor-changing $b~{\to}~s$ quark transition in $B$-meson decays.
Explanations of this anomaly led to development of models either 
with physics beyond the standard model~\cite{nptheory}, 
new weak dynamics~\cite{smtheory}, or strong dynamics~\cite{qcdtheory}. 

A vector-vector $B$-meson decay, such as $B\to \varphi K^{*}$,
is characterized by three complex helicity amplitudes $A_{1\lambda}$ 
which correspond to helicity states $\lambda=-1,0,+1$ of the 
vector mesons. The $A_{10}$ amplitude is expected to dominate~\cite{bvv1}
due to the $(V-A)$ nature of the weak interactions and helicity conservation
in the strong interactions. Experimental results suggest that 
$A_{1+1}$ and $A_{1-1}$ comprise about 50\% 
of the total decay amplitude in 
$B\to\varphi K^{*}$~\cite{babar:vv, belle:phikst}.
Recently, the $\babar$ experiment extended the study  
of the $B^0\to\varphi K^{*0}$ decay to resolve the discrete ambiguity
between the $A_{1+1}$ and $A_{1-1}$ amplitudes~\cite{babar:vt}.

We now investigate the polarization puzzle with a full 
amplitude analysis of the $B^\pm\to\varphi K^{*}(892)^\pm$ decay. 
In this paper, we report twelve independent parameters for the 
three $B^+$ and three $B^-$ decay amplitudes, six of which are 
presented for the first time.
Moreover, we use the dependence on the $K\pi$ invariant mass 
of the interference between the $J^P=1^-$ and $0^+$ 
$(K\pi)^{\pm}$ components~\cite{babar:vt, Aston:1987ir, jpsikpi}
to resolve the discrete ambiguity between the
$A_{1 +1}$ and $A_{1 -1}$ helicity amplitudes.


We use a sample of $383.6\pm 4.2$ million $\FourS\to\BB$ events
collected with the \babar\ detector~\cite{babar} at the PEP-II $e^+e^-$ 
asymmetric-energy storage rings. The $e^+e^-$ center-of-mass energy $\sqrt{s}$ 
is equal to $10.58$ GeV.
Momenta of charged particles are measured 
in a tracking system consisting of a silicon vertex tracker with five 
double-sided layers and a 40-layer drift chamber, both within the 1.5-T 
magnetic field of a solenoid. 
Identification of charged particles is provided 
by measurements of the energy loss in the tracking devices and by 
a ring-imaging Cherenkov detector. 
Photons are detected by a CsI(Tl) electromagnetic calorimeter.


The $B^\pm\to\varphi(1020)K^{*\pm}\to(K^+K^-)(K\pi)^\pm$ candidates 
are analyzed with two $(K\pi)^\pm$ final states, $K^0_S\pi^\pm$ and $K^\pm\pi^0$.
The neutral pseudoscalar mesons are reconstructed
in the final states $K^0_S\to\pi^+\pi^-$ and $\pi^0\to\gamma\gamma$.
We define the helicity angle $\theta_i$ as the angle between the direction
of the $K$ or $K^+$ meson from $K^*\to K\pi$ ($\theta_1$) or $\varphi\to K^+K^-$ 
($\theta_2$) and the direction opposite the $B$ in the $K^*$ or $\varphi$
rest frame, and $\Phi$ as the angle between the decay planes of the two 
systems~\cite{babar:vt}. The differential decay width has four complex amplitudes 
$A_{J\lambda}$ which describe two spin states of the $K\pi$ system 
($J=1$ or $0$) and the three helicity states of the $J=1$ state 
($\lambda=0$ or $\pm 1$):
\begin{eqnarray}
\label{eq:helicityfull}
{d^3\Gamma \over d{\cal H}_1 d{\cal H}_2d\Phi} \propto
\left|~\sum_{} 
A_{J\lambda} Y_{J}^{\lambda}({\cal H}_1,\Phi) Y_{1}^{-\!\lambda}(-{\cal H}_2,0)~\right|^2,
\end{eqnarray}
where ${\cal H}_i=\cos\theta_i$ and
$Y_{J}^{\lambda}$ are the spherical harmonics with $J=1$ 
for $K^{*}(892)$ and $J=0$ for $(K\pi)_0^{*}$.
We reparameterize the amplitudes as
$A_{1\pm1}=(A_{1\parallel}\pm A_{1\perp})/\sqrt{2}$.

We identify $B$ meson candidates using two kinematic variables: 
$m_{\rm{ES}} = [{ (s/2 + \mathbf{p}_{\Upsilon} \cdot 
\mathbf{p}_B)^2 / E_{\Upsilon}^2 - \mathbf{p}_B^{\,2} }]^{1/2}$
and $\Delta{E}=(E_{\Upsilon}E_B-\mathbf{p}_{\Upsilon}\cdot\mathbf{p}_B-s/2)/\sqrt{s}$,
where $(E_B,\mathbf{p}_B)$ is the four-momentum of the $B$ candidate,
and $(E_{\Upsilon},\mathbf{p}_{\Upsilon})$ is the $e^+e^-$ initial state four-momentum, 
both in the laboratory frame.
We require $m_{\rm{ES}}>5.25$ GeV and $|\Delta{E}|<0.1$ GeV.
The requirements on the invariant masses are 
$0.75 < m_{K\!\pi} < 1.05$ GeV, $0.99 < m_{K\!\Kbar} < 1.05$ GeV,
$|m_{\pi\pi}-m_{K^0}|< 12$ MeV, and $120 < m_{\gamma\gamma} < 150$ MeV
for the $K^{*\pm}$, $\varphi$, $K^0_S$, and $\pi^0$, respectively.
For the $K^0_S$ candidates, we also require the cosine of the angle
between the flight direction from the interaction point
and momentum direction to be greater than 0.995
and the measured proper decay time greater than five times its
uncertainty.

To reject the dominant $e^+e^-\to$ quark-antiquark 
background, we use the angle $\theta_T$ between the $B$-candidate 
thrust axis and that of the rest of the event, and a Fisher 
discriminant ${\cal F}$~\cite{bigPRD}. Both variables are 
calculated in the center-of-mass frame. The discriminant 
combines the polar angles of the $B$-momentum vector 
and the $B$-candidate thrust axis with respect to the beam axis,
and two moments of the energy flow around the 
$B$-candidate thrust axis~\cite{bigPRD}.

To reduce combinatorial background with low-momentum
$\pi^0$ candidates, we require ${\cal H}_1<0.6$.
When more than one candidate is reconstructed,
which happens in $7\%$ of events with $K_S^0$ and $17\%$ with $\pi^0$, 
we select the one whose $\chi^2$ of the charged-track vertex fit 
combined with $\chi^2$ of the invariant mass consistency of 
the $K^0_S$ or $\pi^0$ candidate, is the lowest. 
We define the $b$-quark flavor sign $Q$ to be opposite 
to the charge of the $B$ meson candidate.


We use an unbinned, extended maximum-likelihood fit~\cite{babar:vv, babar:vt} 
to extract the event yields $n_{j}^k$ and the parameters of the probability 
density function (PDF) ${\cal P}_{j}^k$.
The index $j$ represents three event categories used in our data model:
the signal $B^\pm\to\varphi(K\pi)^\pm$ ($j = 1$), 
a possible background from $B^\pm\to f_0(980)K^{*\pm}$ ($j = 2$), 
and combinatorial background ($j = 3$).
The superscript $k$ corresponds to the value of $Q=\pm$
and allows for a $C\!P$-violating difference between 
the $B^+$ and $B^-$ decay amplitudes ($A$ and $\Abar$).
In the signal category, the yield and asymmetry of the 
$B^\pm\to\varphi K^*(892)^\pm$ mode, $n_{\rm sig}$ and ${\cal A}_{C\!P}$,
and those of the $B^\pm\to\varphi(K\pi)_0^{*\pm}$ mode are parameterized 
by applying the fraction of $\varphi K^*(892)^\pm$ yield, $\mu^k$, to $n_1^k$.
Hence, $n_{\rm sig}=n_1^+\times\mu^+ + n_1^-\times\mu^-$,
${\cal A}_{C\!P}=(n_1^+\times\mu^+ - n_1^-\times\mu^-)/n_{\rm sig}$,
and the $\varphi(K\pi)_0^{*\pm}$ yield is $n_1^+\times(1-\mu^+) + n_1^-\times(1-\mu^-)$.

The likelihood ${\cal L}_i$ for each candidate $i$ is defined as
${\cal L}_i = \sum_{j,k}n_{j}^k\, 
{\cal P}_{j}^k$({\boldmath ${\rm x}_i$};~$\mu^k$,~{\boldmath$\zeta$},~{\boldmath$\xi$}),
where the PDF is formed based on the following set of observables
{\boldmath ${\rm x}_i$}~$=\{{\cal H}_1$, ${\cal H}_2$, $\Phi$, 
$m_{K\!\pi}$, $m_{K\!\Kbar}$, $\Delta E$, $m_{\rm{ES}}$, ${\cal F}$, $Q$\}
and the dependence on $\mu^k$ and polarization parameters {\boldmath$\zeta$} 
is relevant only for the signal PDF ${\cal P}_{1}^k$.
The remaining PDF parameters {\boldmath$\xi$}
are left free to vary in the fit for the combinatorial 
background and are fixed to the values extracted from 
Monte Carlo (MC) simulation~\cite{geant} and calibration 
$B\to\Dbar\pi$ decays for event categories $j = 1$ and $2$.

\begin{figure}[t]
\centerline{
\setlength{\epsfxsize}{0.5\linewidth}\leavevmode\epsfbox{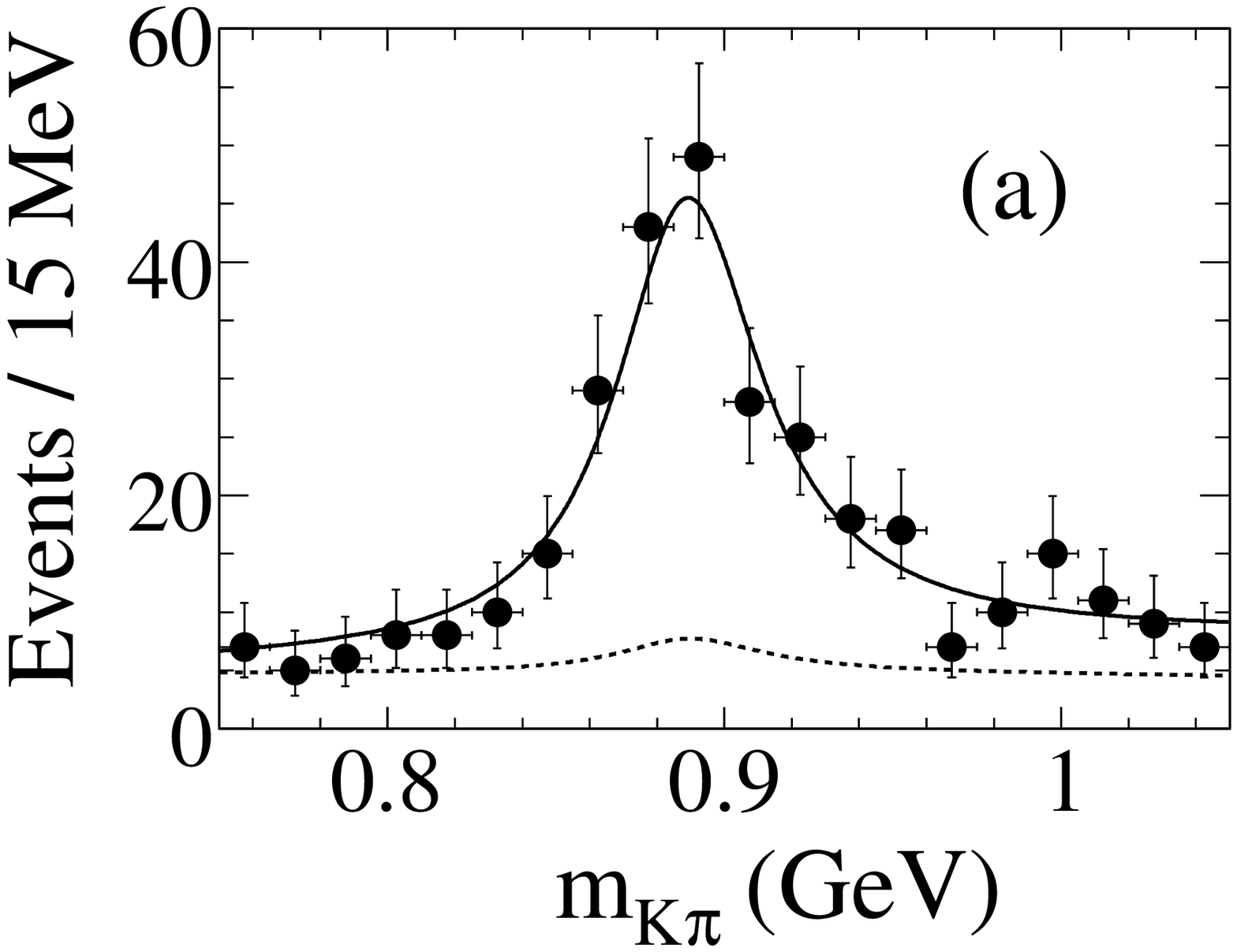}
\setlength{\epsfxsize}{0.5\linewidth}\leavevmode\epsfbox{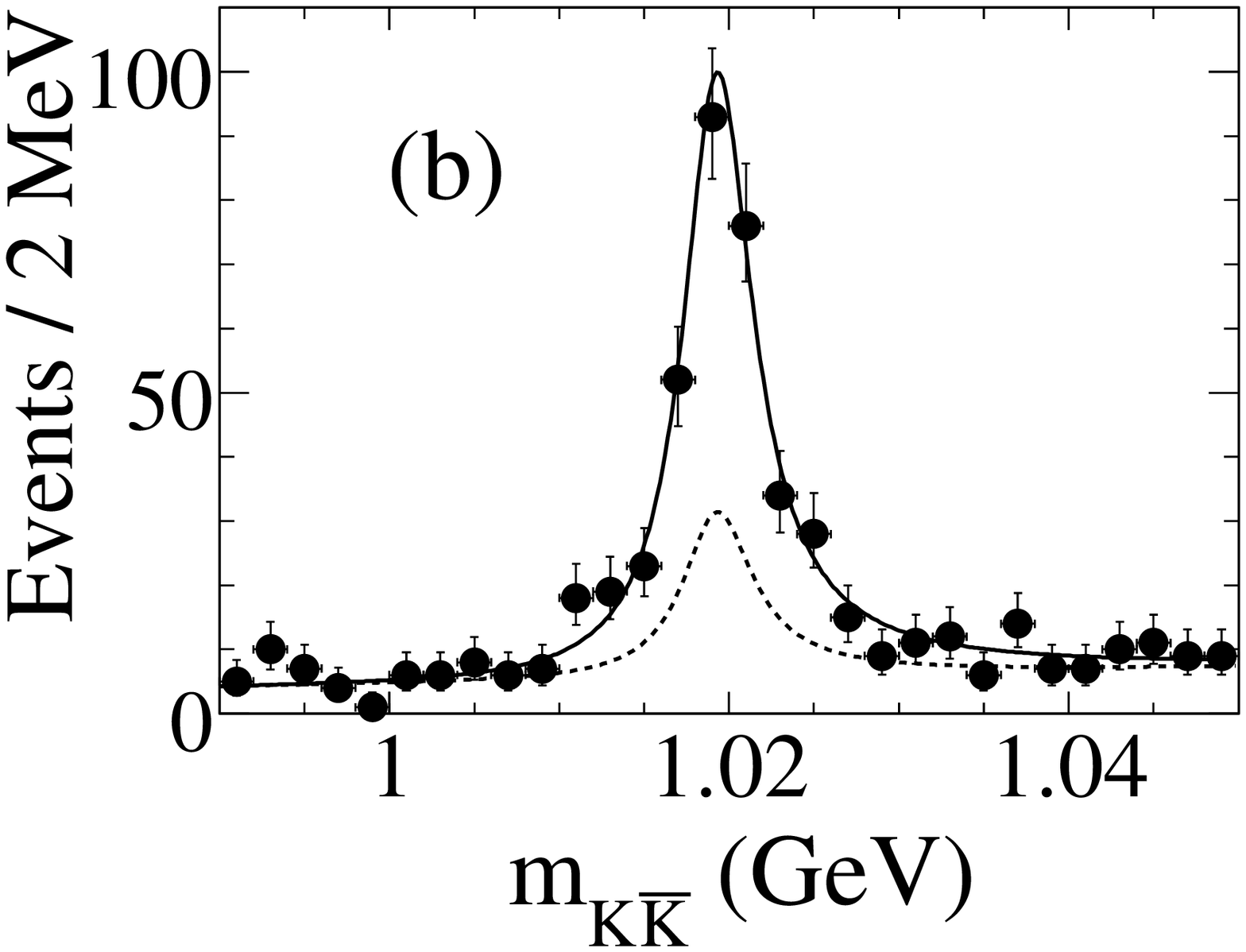}
}
\centerline{
\setlength{\epsfxsize}{0.5\linewidth}\leavevmode\epsfbox{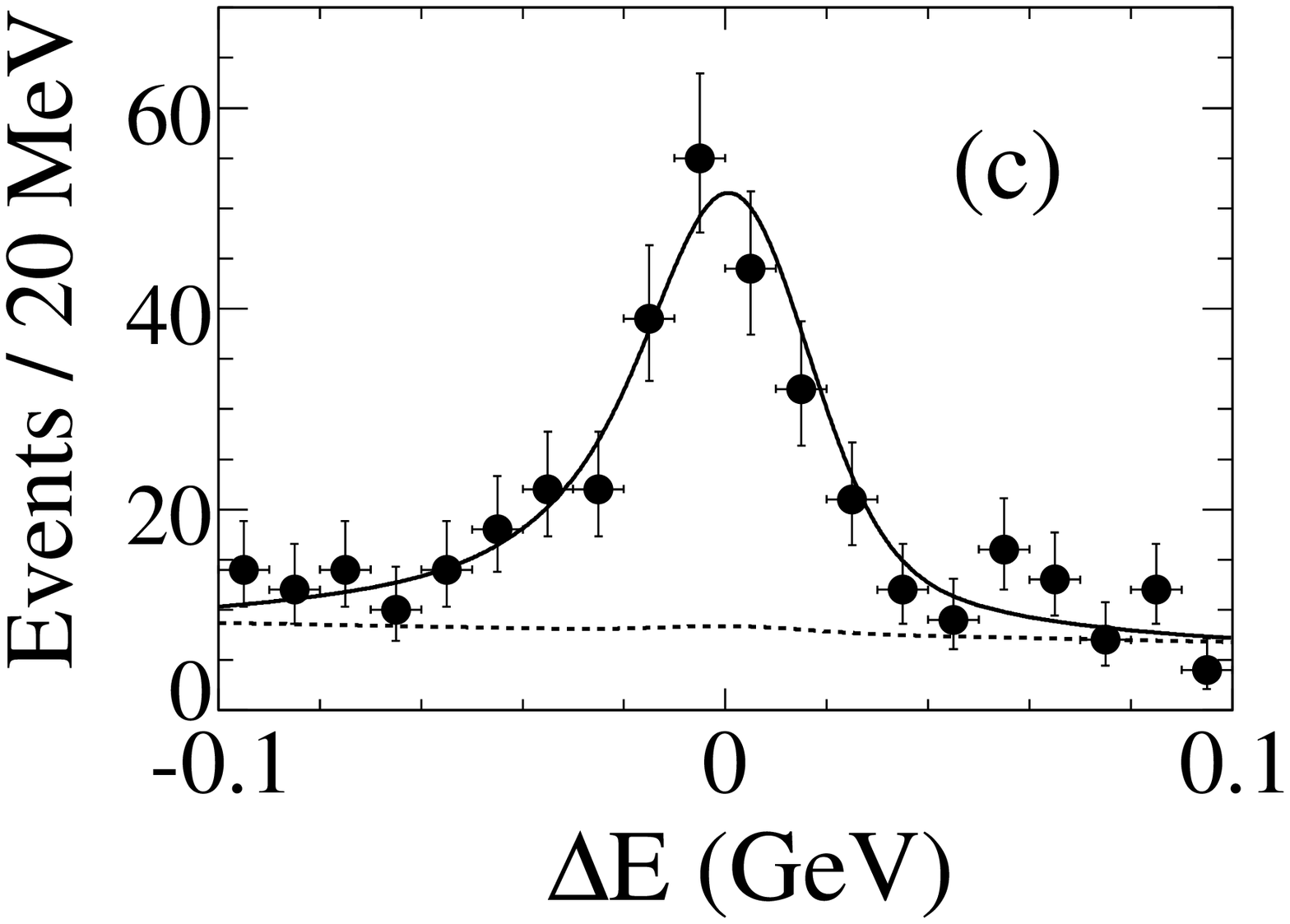}
\setlength{\epsfxsize}{0.5\linewidth}\leavevmode\epsfbox{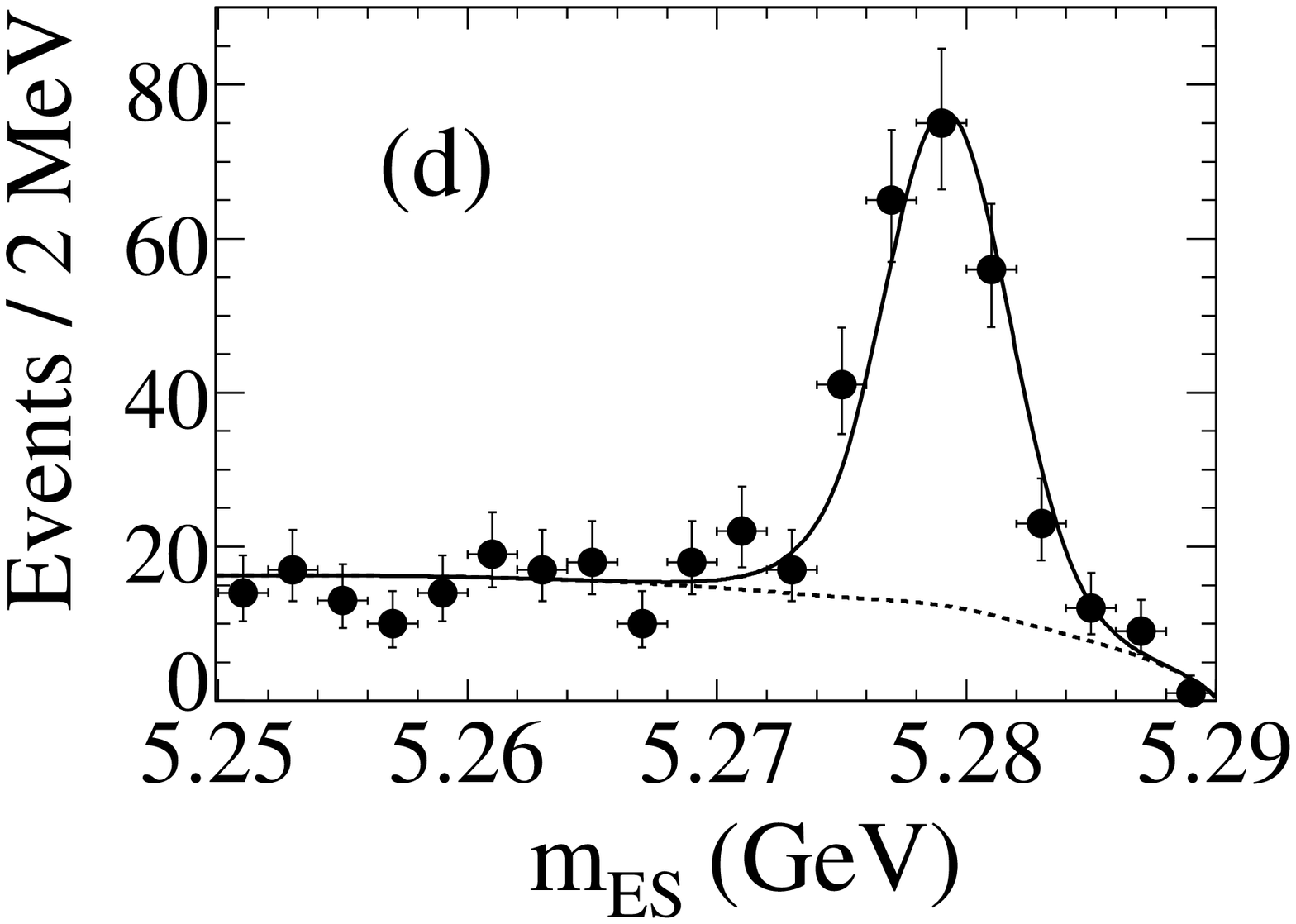}
}
\vspace{-0.3cm}
\caption{\label{fig:projection1} 
Projections onto the variables (a) $m_{K\!\pi}$, (b) $m_{K\!\Kbar}$, (c) $\Delta E$,
and (d) $m_{\rm ES}$ for the signal $B^\pm\to\varphi(K\pi)^\pm$ candidates
with a requirement discussed in the text.
The solid (dashed) lines show the signal-plus-background
(background) PDF projections.
}
\end{figure}
\begin{figure}[t]
\centerline{
\setlength{\epsfxsize}{0.5\linewidth}\leavevmode\epsfbox{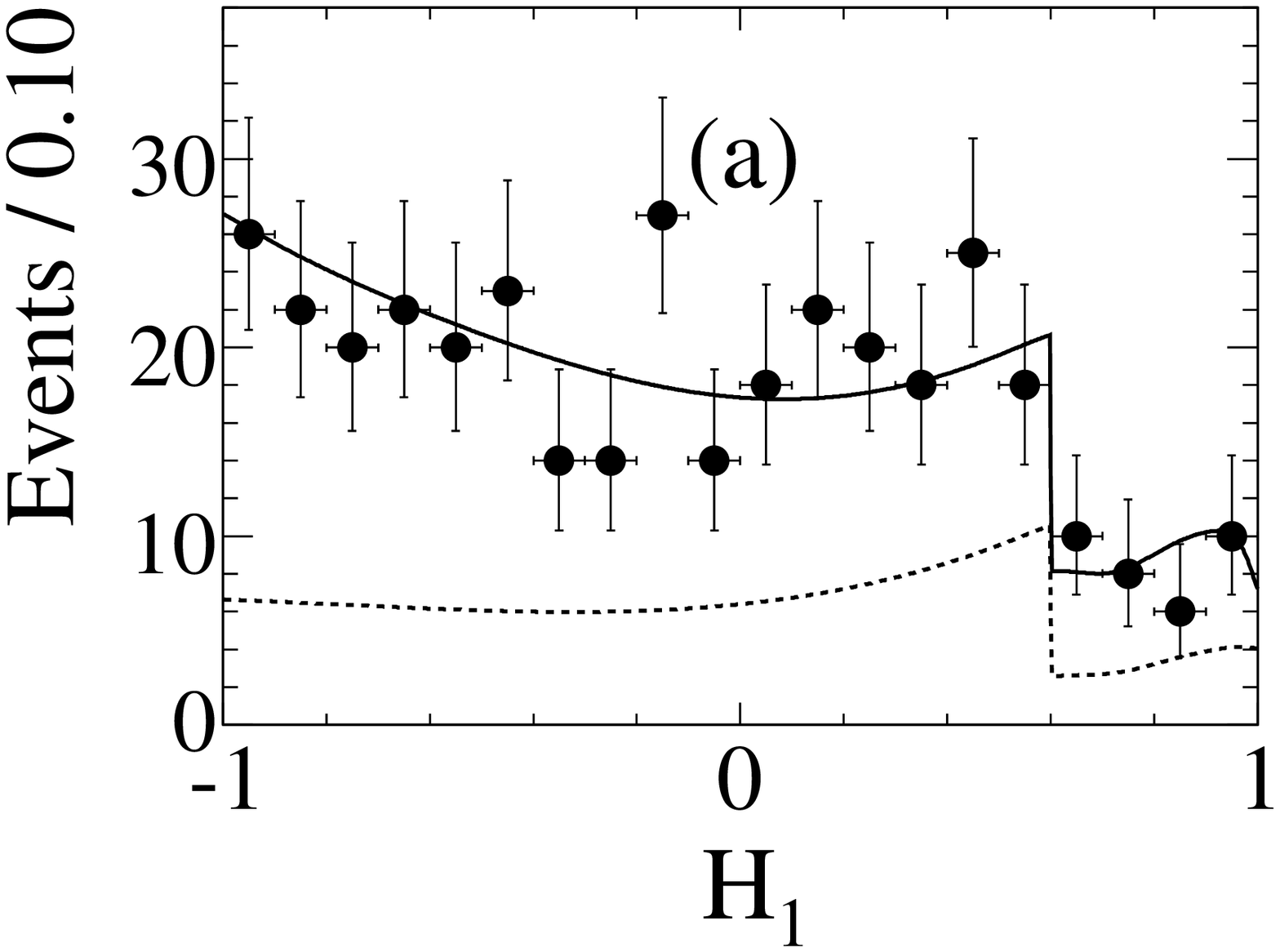}
\setlength{\epsfxsize}{0.5\linewidth}\leavevmode\epsfbox{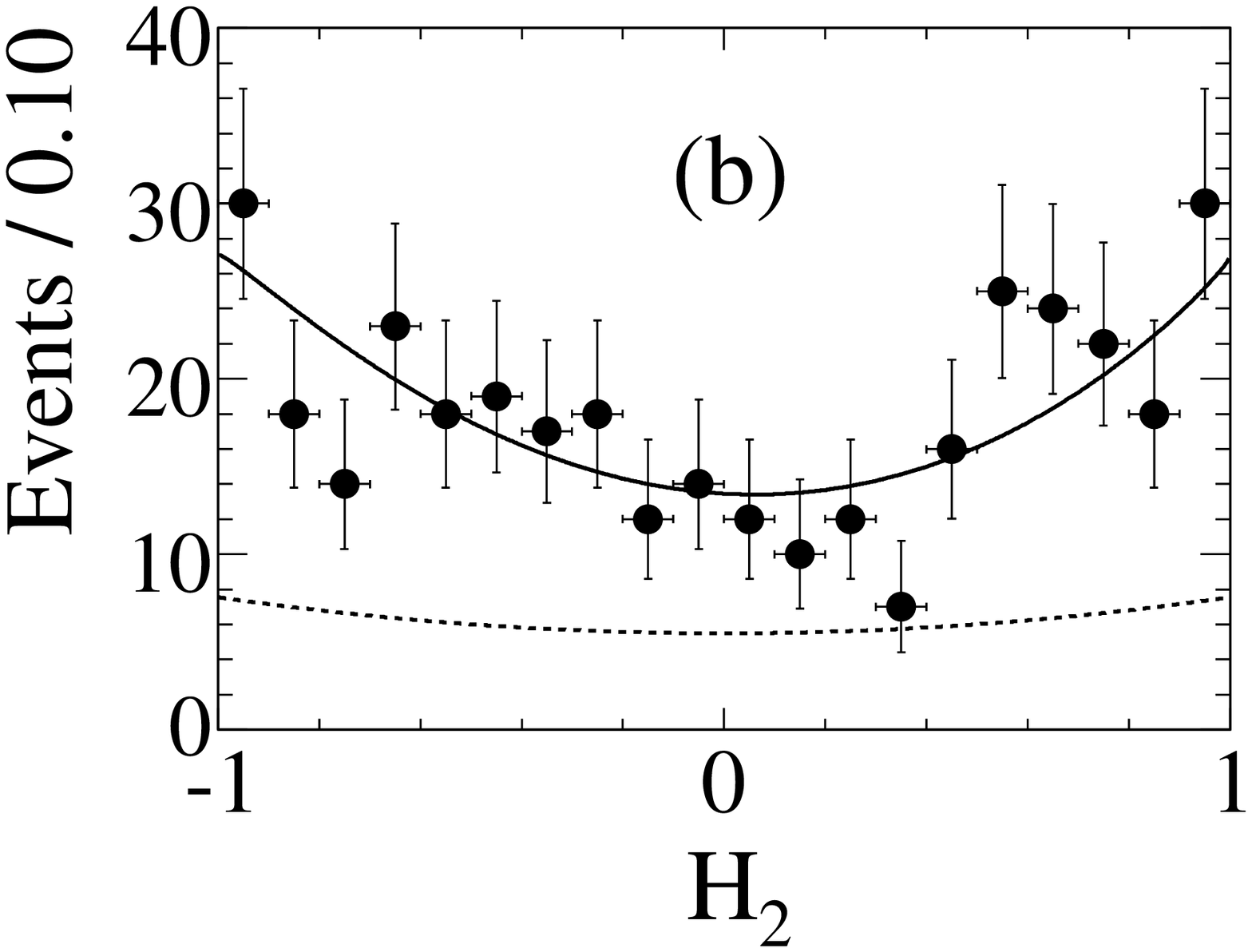}
}
\centerline{
\setlength{\epsfxsize}{0.5\linewidth}\leavevmode\epsfbox{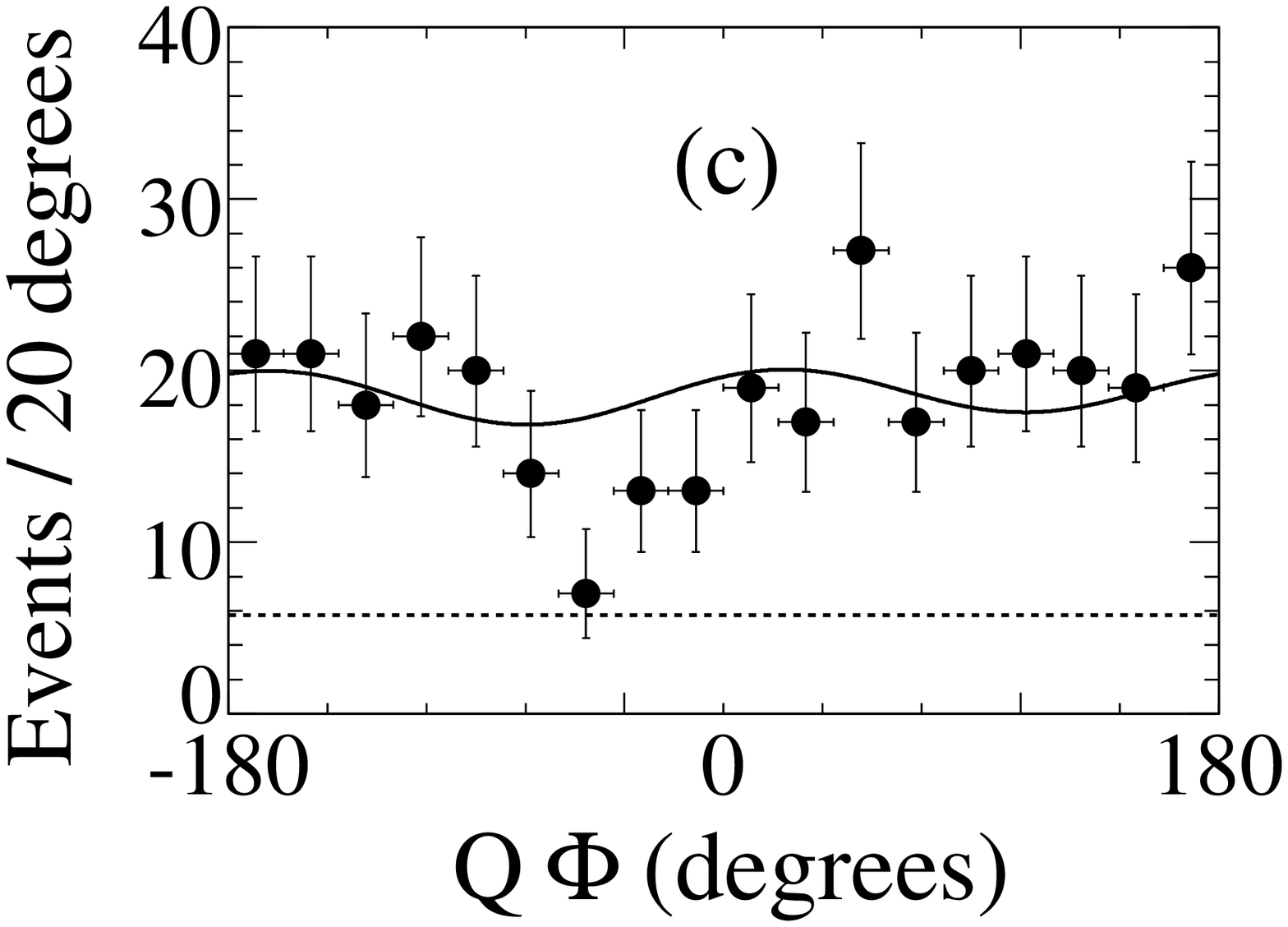}
\setlength{\epsfxsize}{0.5\linewidth}\leavevmode\epsfbox{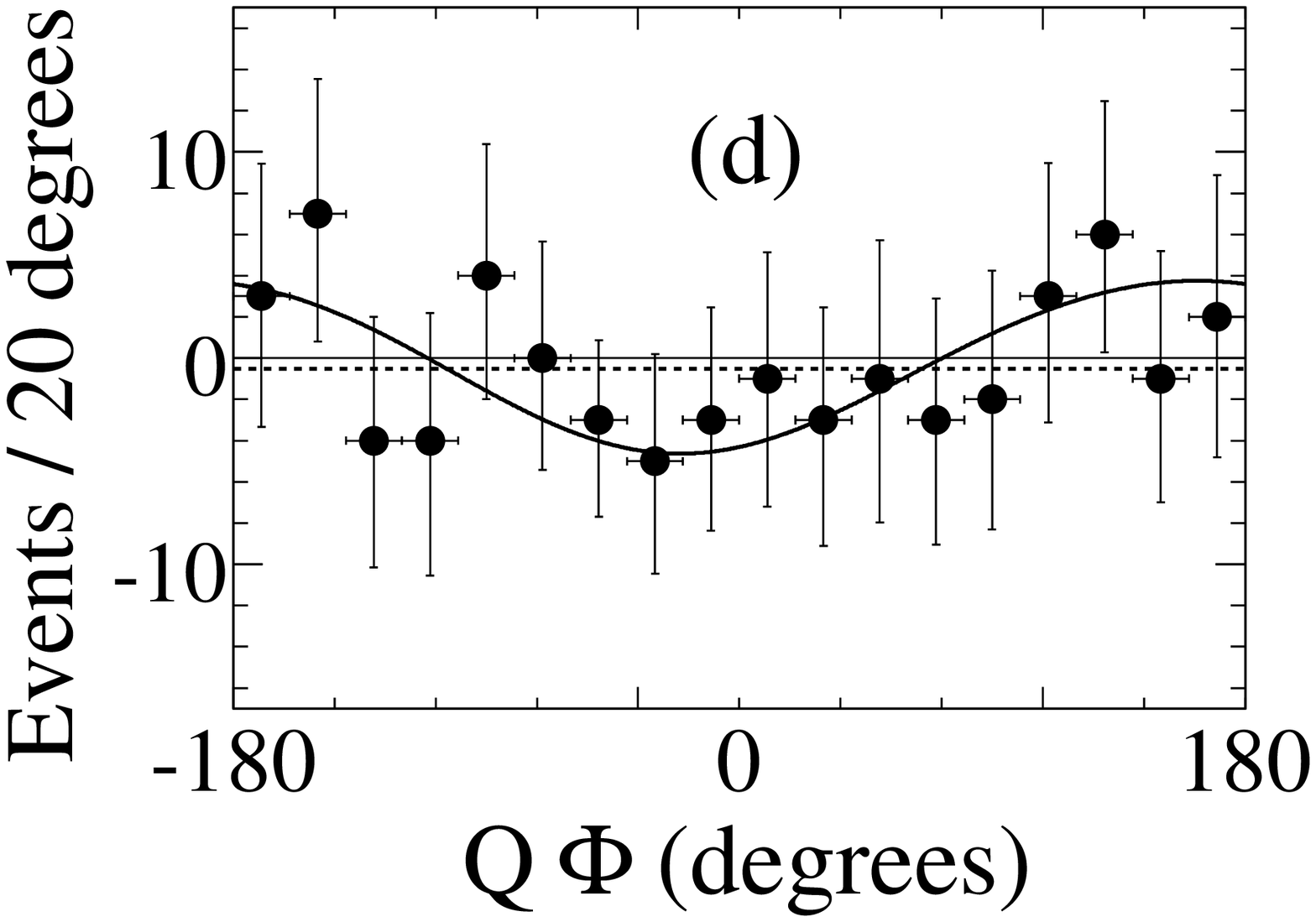}
}
\vspace{-0.3cm}
\caption{\label{fig:projection2} 
Projections onto the variables (a) 
${\cal H}_1$, (b) ${\cal H}_2$, (c) $Q~\!\Phi$, and (d) 
the differences between
the $Q~\!\Phi$ projections for events with 
${\cal H}_1~\!{\cal H}_2>0$ and with
${\cal H}_1~\!{\cal H}_2<0$ 
for the signal $B^\pm\to\varphi(K\pi)^\pm$  candidates
following the solid (dashed) line definitions in Fig.~\ref{fig:projection1}.
The step in the ${\cal H}_1$ PDF distributions is due to the
selection requirement  ${\cal H}_1<0.6$ in the $B^\pm\to\varphi(K^\pm\pi^0)$
channel.
}
\end{figure}

The helicity part of the signal PDF is the 
ideal angular distribution from Eq.~(\ref{eq:helicityfull}),
multiplied by an empirical acceptance function
${\cal{G}}({\cal H}_1,{\cal H}_2,\Phi)
\equiv{\cal{G}}_1({\cal H}_1)\times{\cal{G}}_2({\cal H}_2)$.
Here, the amplitudes $A_{J\lambda}$ are expressed in terms of
the polarization parameters 
{\boldmath$\zeta$}~$\equiv\{f_L$, $f_{\perp}$, $\phi_{\parallel}$, 
$\phi_{\perp}$, $\delta_0$, ${\cal A}_{C\!P}^0$, ${\cal A}_{C\!P}^{\perp}$, 
$\Delta \phi_{\parallel}$, $\Delta \phi_{\perp}$, $\Delta\delta_0$\}
defined in Table~\ref{tab:results}.
$C\!P$-violating differences are incorporated via the replacements 
in Eq.~(\ref{eq:helicityfull}) for $B^{+}$ decays:
$f_L\to f_L\times(1+{\cal A}^0_{C\!P}\times Q)$, 
$f_\perp\to f_\perp\times(1+{\cal A}^\perp_{C\!P}\times Q)$, 
$\phi_\parallel\to(\phi_\parallel+\Delta\phi_\parallel\times Q)$, 
$\phi_\perp\to(\phi_\perp+{\pi/2}+(\Delta\phi_\perp+{\pi/2})\times Q)$,
and $\delta_0\to(\delta_0+\Delta\delta_0\times Q)$.

A relativistic spin-$J$ Breit--Wigner amplitude parameterization 
is used for the resonance masses~\cite{pdg2006,f0mass}, and the 
$(K\pi)^{*\pm}_0$ $m_{K\!\pi}$ amplitude is parameterized with 
the LASS function~\cite{Aston:1987ir}.
The latter includes the $K_0^{*}(1430)^\pm$ resonance
together with a nonresonant component.
The interference between the $J=0$ and $1$ $(K\pi)^{\pm}$ contributions 
is modeled with the three terms $2{\cal R\rm e}(A_{1\lambda} A^*_{00})$ 
in Eq.~(\ref{eq:helicityfull}) with the four-dimensional angular and $m_{K\!\pi}$ 
parameterization and with dependence on $\mu^k$ and {\boldmath$\zeta$}.

\begingroup
\begin{table*}[t]
\caption{\label{tab:results}
Summary of results for the $B^\pm\to\varphi K^{*}(892)^\pm$ decay.
The twelve primary results are presented for the two decay
subchannels along with the combined results, where the branching
fraction ${\cal B}$ is computed using the number of signal events 
$n_{\rm sig}$ and the total selection efficiency $\varepsilon$, 
which includes the daughter branching fractions~\cite{pdg2006} and
the reconstruction efficiency $\varepsilon_{\rm reco}$ obtained
from MC simulation.
The definition of the six $C\!P$-violating parameters allows for
differences between the $B^+$ and ${B}^-$ decay amplitudes
$A$ and $\Abar$ with superscript $Q=-$ and $+$, respectively.
The systematic uncertainties are quoted last and are not included 
for the intermediate primary results in each subchannel.
The dominant fit correlation coefficients (${\cal C}$) are presented, where
we show correlations of ${\delta_0}$  with ${\phi_\parallel}/{\phi_\perp}$
and of ${\Delta\delta_0}$ with ${\Delta\phi_\parallel}/{\Delta\phi_\perp}$.
}
\begin{center}
{
\begin{ruledtabular}
\setlength{\extrarowheight}{1.5pt}
\begin{tabular}{cccccc}
\vspace{-3mm} & & \\
   parameter
 & definition
 & $K^{*}(892)^\pm\to K^0_S\pi^\pm$ 
 & $K^{*}(892)^\pm\to K^\pm\pi^0$  
 & combined
 & ${\cal C}$ 
\cr
\vspace{-3mm} & & & \\
\hline
\vspace{-3mm} & & & \\
  ${\cal B}$ 
 & $\Gamma/\Gamma_{\rm total}$ 
 & $(10.5\pm1.4)\times10^{-6}$
 & $(11.6\pm1.5)\times10^{-6}$
 & $(11.2\pm1.0\pm0.9)\times10^{-6}$
\cr
\vspace{-3mm} & & & \\
  ${f_L}$  
 & ${|A_{10}|^2/\Sigma|A_{1\lambda}|^2}$
 & $0.51\pm0.07$
 & $0.46^{+0.10}_{-0.09}$
 & $0.49\pm0.05\pm 0.03$ 
 & \multirow{2}{13mm}{~{\Large\}}$-58\%$}
\cr
\vspace{-3mm} & & & \\
  ${f_\perp}$ 
 & ${|A_{1\perp}|^2/\Sigma|A_{1\lambda}|^2}$
 & $0.22^{+0.07}_{-0.06}$
 & $0.21^{+0.09}_{-0.08}$
 & $0.21\pm 0.05\pm 0.02$
 &
\cr
\vspace{-3mm} & & & \\
  ${\phi_\parallel}-\pi$ 
 & ${\rm arg}(A_{1\parallel}/A_{10})-\pi$
 & $-0.75^{+0.28}_{-0.24}$
 & $-0.77\pm0.35$
 & $-0.67\pm{0.20}\pm0.07$
 &  \multirow{2}{13mm}{~{\Large\}}~$+56\%$}
\cr
\vspace{-3mm} & & & \\
  ${\phi_\perp}-\pi$  
 & ${\rm arg}(A_{1\perp}/A_{10})-\pi$
 & $-0.15\pm 0.24$
 & $-0.89^{+0.40}_{-0.46}$
 & $-0.45\pm{0.20}\pm0.03$
 &  
\cr
\vspace{-3mm} & & & \\
  ${\delta_0}-\pi$  
 & ${\rm arg}(A_{00}/A_{10})-\pi$
 & $-0.25\pm0.24$
 & $+0.11\pm{0.31}$
 & $-0.07\pm0.18\pm0.06$
 &                        {$+37\%/+36\%$}
\cr
\vspace{0mm} & & & \\
  ${\cal A}_{C\!P}$ 
 & $(\Gamma^+-\Gamma^-)/(\Gamma^++\Gamma^-)$
 & $-0.09\pm0.13$
 & $+0.07\pm0.13$
 & $0.00\pm0.09\pm0.04$
 &
\cr
\vspace{-3mm} & & & \\
  ${\cal A}_{C\!P}^0$ 
 & $(f_L^+-f_L^-)/(f_L^++f_L^-)$
 & $+0.24\pm0.15$
 & $+0.09\pm0.20$
 & $+0.17\pm0.11\pm0.02$ 
 &  \multirow{2}{13mm}{~{\Large\}}$-50\%$}
\cr
\vspace{-3mm} & & & \\
  ${\cal A}_{C\!P}^{\perp}$ 
 & $(f_\perp^+-f_\perp^-)/(f_\perp^++f_\perp^-)$
 & $+0.12\pm{0.31}$
 & $+0.41^{+0.54}_{-0.40}$
 & $+0.22\pm{0.24}\pm0.08$
 &  
\cr
\vspace{-3mm} & & & \\
  $\Delta \phi_{\parallel}$  
 & $(\phi_{\parallel}^+-\phi_{\parallel}^-)/2$ 
 & $+0.02\pm0.28$
 & $+0.22\pm{0.35}$
 & $+0.07\pm0.20\pm0.05$
 &  \multirow{2}{13mm}{~{\Large\}}~$+57\%$}
\cr
\vspace{-3mm} & & & \\
  $\Delta \phi_{\perp}$  
 & $(\phi_{\perp}^+-\phi_{\perp}^--\pi)/2$ 
 & $+0.18\pm 0.24$
 & $+0.48^{+0.46}_{-0.40}$
 & $+0.19\pm{0.20}\pm0.07$
 & 
\cr
\vspace{-3mm} & & & \\
  ${\Delta\delta_0}$  
 & $(\delta_0^+-\delta_0^-)/2$ 
 & $+0.13\pm{0.24}$
 & $+0.34\pm0.31$
 & $+0.20\pm0.18\pm0.03$ 
 &                        {$+37\%/+37\%$}
\cr
\vspace{0mm} & & & \\
   $n_{\rm sig}$ 
 &
 & $102\pm 13\pm 6$  
 & $117^{+15}_{-16}\pm 7$  
 & 
 & 
\cr
\vspace{-3mm} & & & \\
    $\varepsilon$  
 &
 & $(2.53\pm0.13)$ \%
 & $(2.59\pm0.17)$ \%
 & 
 & 
\cr 
\vspace{-3mm} & & & \\
   $\varepsilon_{\rm reco}$  
 &
 & $(22.3\pm1.2)$ \%
 & $(16.0\pm1.0)$ \%
 & 
 & 
\cr
\vspace{-3mm} & & & \\
\end{tabular}
\end{ruledtabular}
}
\end{center}
\end{table*}
\endgroup

The signal PDF 
for a given candidate $i$ is a joint PDF for the helicity angles
and resonance mass as discussed above, and the product of 
the PDFs for each of the remaining variables.
The combinatorial background PDF is the product of the 
PDFs for independent variables and is found to describe 
well both the dominant quark-antiquark background and the 
background from random combinations of $B$ tracks.
The signal and background PDFs are illustrated in
Figs.~\ref{fig:projection1} and~\ref{fig:projection2}.
For illustration,
the signal fraction is enhanced with a requirement on the 
signal-to-background probability ratio, calculated with 
the plotted variable excluded, that is at least 50\% 
efficient for signal $B^\pm\to\varphi(K\pi)^\pm$ events.
We use a sum of Gaussian functions 
for the parameterization of the signal PDFs 
for $\Delta E$, $m_{\rm{ES}}$, and ${\cal F}$.
For the combinatorial background, we use polynomials,
except for $m_{\rm{ES}}$ and ${\cal F}$ distributions
which are parameterized by an empirical phase-space 
function and by Gaussian functions, respectively.
Resonance production occurs in the background and 
is taken into account in the PDF.


We observe a nonzero $B^\pm\to\varphi K^{*}(892)^\pm$
yield with significance, including systematic uncertainties,
of more than 10$\sigma$.
The significance is defined as the square root of the change in 
$2\ln{\cal L}$ when the yield is constrained to zero in the 
likelihood ${\cal L}$.
In Table~\ref{tab:results}, results of the fit are presented,
where the combined results are obtained from the simultaneous fit
to the two decay subchannels.


We repeat the fit by varying the fixed parameters 
in {\boldmath$\xi$} within their uncertainties 
and obtain the associated systematic uncertainties.
We allow for a flavor-dependent acceptance function
and reconstruction efficiency in the study of asymmetries.
The biases from the finite resolution of the angle measurements, 
the dilution due to the presence of fake combinations, 
or other imperfections in the signal PDF model are estimated 
with MC simulation.

The nonresonant $K^+K^-$ contribution under the 
$\varphi$ is accounted for with the $B^0\to f_0 K^{*0}$ category.
Its yield is consistent with zero. 
The $m_{K\!\Kbar}$ PDF shape in this category
is varied from the resonant to phase-space
and the yield is varied from the observed value to the extrapolation from the 
neutral $B$-decay mode~\cite{babar:vt} to estimate the systematic uncertainties.
Additional systematic uncertainty originates  
from other potential $B$ backgrounds, 
which we estimate can contribute 
at most a few events to the signal component.
The systematic uncertainties in efficiencies are dominated 
by those in particle identification, track finding,
and $K^0_S$ and $\pi^0$ selection.
Other systematic effects arise from event-selection criteria, 
$\varphi$ and $K^{*0}$ branching fractions, and the number of $B$ mesons.


The yield of the $\varphi({K\pi})^{*\pm}_0$ contribution 
is $57^{+14}_{-13}$ events 
with a statistical significance of  {7.9}$\sigma$,
combining the $|A_{00}|^2$ term and the interference terms
$2{\cal R\rm e}(A_{1\lambda} A^*_{00})$, which confirms the 
significant $S$-wave $K\pi$ contribution observed in
the neutral $B$-decay mode~\cite{babar:vt}.
The dependence of the interference on the $K\pi$ invariant 
mass~\cite{babar:vt, Aston:1987ir, jpsikpi} allows us to reject 
the other solution near ($2\pi-\phi_{\parallel},\pi-\phi_{\perp}$)
relative to that in Table~\ref{tab:results}
with significance of {6.3}$\sigma$, including systematic uncertainties.

The $(V-A)$ structure of the weak interactions,
helicity conservation in strong interactions,
and the $s$-quark spin flip suppression in the penguin 
decay diagram suggest $|A_{10}|\gg|A_{1+1}|\gg|A_{1-1}|$~\cite{bvv1}. 
This expectation disagrees with our observed value of $f_L$.
We obtain the solution 
${\phi_\parallel}\simeq{\phi_\perp}$ without discrete ambiguities,
which is consistent with the approximate decay amplitude hierarchy 
$|A_{10}|\simeq|A_{1+1}|\gg|A_{1-1}|$.

We find that ${\phi_\perp}$ and ${\phi_\parallel}$ deviate from either 
$\pi$ or zero by more than {3.1}$\sigma$ and {2.4}$\sigma$, respectively, 
including systematic uncertainties. This indicates the presence of 
final-state interactions not accounted for in naive factorization.
Our measurements of the six $C\!P$-violating parameters are 
consistent with zero and exclude a significant part of 
the physical region. We find no evidence of $C\!P$ violation 
in this decay.

In summary, we have performed a full amplitude analysis and searched 
for $C\!P$-violation in the angular distribution of the 
$B^\pm\to\varphi K^{*\pm}$ decay.
Our results are summarized in Table~\ref{tab:results} and
supersede our prior measurements in Ref.~\cite{babar:vv}.
These results find substantial $A_{1+1}$ amplitude in the 
$B^\pm\to\varphi K^{*\pm}$ decay and point to physics outside 
the standard model or new dynamics~\cite{nptheory,smtheory,qcdtheory}.


We are grateful for the excellent luminosity and machine conditions
provided by our \pep2\ colleagues,
and for the substantial dedicated effort from
the computing organizations that support \babar.
The collaborating institutions wish to thank
SLAC for its support and kind hospitality.
This work is supported by
DOE
and NSF (USA),
NSERC (Canada),
CEA and
CNRS-IN2P3
(France),
BMBF and DFG
(Germany),
INFN (Italy),
FOM (The Netherlands),
NFR (Norway),
MIST (Russia),
MEC (Spain), and
PPARC (United Kingdom).
Individuals have received support from the
Marie Curie EIF (European Union) and
the A.~P.~Sloan Foundation.


\bibliographystyle{h-physrev2-original}   

\begin{thebibliography}{99}

\bibitem{babar:vv}
$\babar$ Collaboration, B.~Aubert {\it et al.},
Phys.\ Rev.\ Lett.\ {\bf 91}, 171802 (2003);
{\bf 93}, 231804 (2004).

\bibitem{belle:phikst}
Belle Collaboration, K.-F. Chen {\it et al.},
Phys.\ Rev.\ Lett.\ {\bf 91}, 201801 (2003);
{\bf 94}, 221804 (2005).

\bibitem{belle:rhokst}
Belle Collaboration, J. Zhang {\it et al.},
Phys.\ Rev.\ Lett.\ {\bf 95}, 141801 (2005).

\bibitem{babar:rhokst}
$\babar$ Collaboration, B.~Aubert {\it et al.},
Phys.\ Rev.\ Lett.\ {\bf 97}, 201801 (2006).

\bibitem{babar:vt}
$\babar$ Collaboration, B.~Aubert {\it et al.},
Phys.\ Rev.\ Lett.\ {\bf 98}, 051801 (2007);
arXiv:0705.0398 [hep-ex].

\bibitem{bvvreview2006}
A.~V.~Gritsan and J.~G.~Smith, ``Polarization in $B$ Decays''
review in~\cite{pdg2006}, J. Phys. G33, 833 (2006).

\bibitem{pdg2006}
Particle Data Group,  W.-M. Yao {\it et al.}, J. Phys. G33, 1 (2006).

\bibitem{nptheory}
Y.~Grossman,  Int.\ J.\ Mod.\ Phys.\ A {\bf 19}, 907 (2004);
E.~Alvarez {\it et al.}, Phys.\ Rev.\ D {\bf 70}, 115014 (2004);
P.~K.~Das and K.~C.~Yang, Phys.\ Rev.\ D {\bf 71}, 094002 (2005);
C.~H.~Chen and C.~Q.~Geng, Phys.\ Rev.\ D {\bf 71}, 115004 (2005);
Y.~D.~Yang {\it et al.}, Phys.\ Rev.\ D {\bf 72}, 015009 (2005);
K.~C.~Yang, Phys.\ Rev.\ D {\bf 72}, 034009 (2005);
S.~Baek, Phys.\ Rev.\ D {\bf 72}, 094008 (2005);
C.~S.~Huang {\it et al.}, Phys.\ Rev.\ D {\bf 73}, 034026 (2006);
C.~H.~Chen and H.~Hatanaka, Phys.\ Rev.\ D {\bf 73}, 075003 (2006);
A.~Faessler {\it et al.},  Phys.\ Rev.\ D {\bf 75}, 074029 (2007).

\bibitem{smtheory}
A.~L.~Kagan, Phys.\ Lett.\ B {\bf 601}, 151 (2004);
H.~n.~Li and S.~Mishima, Phys.\ Rev.\ D {\bf 71}, 054025 (2005);
C.-H. Chen et al., Phys.\ Rev.\ D {\bf 72}, 054011 (2005);
M.~Beneke {\it et al.}, Phys.\ Rev.\ Lett.\  {\bf 96}, 141801 (2006),
arXiv:hep-ph/0612290;
C.-H. Chen and C.-Q. Geng,
Phys.\ Rev.\ D {\bf 75}, 054010 (2007).

\bibitem{qcdtheory}
C.~W.~Bauer {\it et al.}, Phys.\ Rev.\ D {\bf 70}, 054015 (2004);
P.~Colangelo {\it et al.}, Phys.\ Lett.\ B {\bf 597}, 291 (2004);
M.~Ladisa {\it et al.}, Phys.\ Rev.\ D {\bf 70}, 114025 (2004);
H.~Y.~Cheng {\it et al.}, Phys.\ Rev.\ D {\bf 71}, 014030 (2005).

\bibitem{bvv1}
A.~Ali {\it et al.}, Z.\ Phys.\ C {\bf 1}, 269 (1979); 
G.~Valencia, Phys.\ Rev.\ D {\bf 39}, 3339 (1989);
G. Kramer and W.F. Palmer, Phys.\ Rev.\ D {\bf 45}, 193 (1992);
H.-Y.~Cheng and K.-C.~Yang, Phys.\ Lett.\ B {\bf 511}, 40 (2001);
C.-H. Chen {\it et al.}, Phys.\ Rev.\ D {\bf 66}, 054013 (2002);
M.~Suzuki, Phys.\ Rev.\ D {\bf 66}, 054018 (2002);
A.~Datta and D.~London, Int.\ J.\ Mod.\ Phys.\ A {\bf 19}, 2505 (2004).

\bibitem{Aston:1987ir}
LASS Collaboration,  D.~Aston {\it et al.},
Nucl.\ Phys.\ B {\bf 296}, 493 (1988);
W.~M.~Dunwoodie, private communications.

\bibitem{jpsikpi}
$\babar$ Collaboration, B.~Aubert {\it et al.},
Phys.\ Rev.\ D {\bf 71}, 032005 (2005);
Phys.\ Rev.\ D {\bf 72}, 072003 (2005).

\bibitem{babar}
\babar\ Collaboration, B.~Aubert {\it et al.},
Nucl.\ Instrum.\ Methods {\bf A479}, 1 (2002).

\bibitem{bigPRD}
$\babar$ Collaboration, B.~Aubert {\it et al.},
Phys.\ Rev.\ D {\bf 70}, 032006 (2004).

\bibitem{f0mass}
E791 Collaboration, E. M. Aitala {\it et al.},
Phys. Rev. Lett. {\bf 86}, 765 (2001).

\bibitem{geant} S.~Agostinelli {\it et al.},
{Nucl.\ Instr.\ Meth.\xspace} A {\bf 506}, 250 (2003).

\end{thebibliography}

\end{document}